# Autoantibody recognition mechanisms of p53 epitopes

J. C. Phillips

Dept. of Physics and Astronomy, Rutgers University, Piscataway, N. J., 08854

Abstract

There is an urgent need for economical blood-based, noninvasive molecular biomarkers to assist in the detection and diagnosis of cancers in a cost-effective manner at an early stage, when curative interventions are still possible. Serum autoantibodies are attractive biomarkers for early cancer detection, but their development has been hindered by the punctuated genetic nature of the ten million known cancer mutations. A recent study of 50,000 patients (Pedersen et al., 2013) showed p53 15-mer epitopes are much more sensitive colon cancer biomarkers than p53, which in turn is a more sensitive cancer biomarker than any other protein. The function of p53 as a nearly universal "tumor suppressor" is well established, because of its strong immunogenicity in terms of not only antibody recruitment, but also stimulation of autoantibodies. Here we examine bioinformatic fractal scaling analysis for identifying sensitive epitopes from the p53 amino acid sequence, and show how it could be used for early cancer detection (ECD). We trim 15-mers to 7-mers, and identify specific 7-mers from other species that could be more sensitive to aggressive human cancers, such as liver cancer.

**1. Introduction** In general protein-protein binding occurs at "hot spots" which are usually enriched in tryptophan, tyrosine and arginine, and hydrophobic occlusion of solvent is found to be a necessary condition for strong binding [1,2]. Interest in specific molecular biomarkers for early cancer detection is growing because of evidence that suggests that autoantibodies stimulated by cancer cells may share specific paratopes that selectively bind to p53 epitopes [3-7]. The superior selectivity of p53 epitopes for autoantibody paratopes [8] suggests that more specific mechanisms may be involved, such as interactions with paratope aromatic side chains and hydrophilic residues [2,9]. Among all proteins p53 is much more hydrophilic than average, and it is also elastically much softer, with about half its structure in its center dominated by β strands [10,11], while the remainder (especially the N-terminal quarter) is disordered. Because we are focused only on p53 and its epitopes, we are able to avoid the enormous complexity of millions of cancer genomic mutations [12]. Multiple signaling pathways, and/or multiple driver gene mutations, may be reflected simply in different p53 epitopic response patterns.

The two bioinformatic scales reported here are the modern thermodynamically second order MZ hydropathicity scale, based on protein conformational self-organized criticality of 5000 protein segments [13,14], and the β strand exposed residue amino acid propensity scale, based on a survey of nearly 2000 β strand structures by FTI [15]. Parallel calculations carried out with the thermodynamically first order KD hydropathicity scale, based on water-air protein unfolding [16], and the FTI β strand buried residue amino acid propensity scale [15], gave weaker results and are not reported, except for one example.

**2. Materials and Methods** How does p53 function as a universal tumor suppressor, the subject of nearly $10^5$ papers? It acts as transcriptional activator, controlling the expression of a variety of genes important in cell cycle regulation and apoptosis [17]. Normally conformational changes in globular proteins, either evolutionary or mutational in origin, are best described with hydropathic scales Ψ, with β strand propensity a secondary physico-chemical factor [18-20]. Because we are interested in epitopes with sizes between 7- and 15-mers, we display in Fig. 1 Ψ(n,9) and Ψ(n,13), where Ψ(n) is the MZ hydropathicity [13] of the nth amino acid in human p53, and superscale Ψ(n,W) is Ψ(n) averaged over a sliding window of wave length W.

While Ψ(n) fluctuates rapidly with n, superscale Ψ(n,W) is smoothed and enables us to examine superscaled profiles. As expected from earlier superscaled profiles of other proteins, including amyloids, superscaled profiles display several features important for protein folding and aggregation, including a DNA binding region hydrophobic plateau, and novel synchronized level elastic hinges [20] (see figure caption). These long-range level double hinges are an example of evolutionary optimization of long-range interactions not included in most discussions of epitope-paratope contact interactions [9]. The best value of W for superscaling depends on the property of interest, here the epitopes studied in [8].

In Fig. 2 we compare the superscale profile of FTI β(n,9) with MZ ψ(n,9), and are pleasantly surprised. The central DNA binding region 102-292 plateau, which already had abrupt edges in Fig. 1 for Ψ(n,9), has even more abrupt edges for β(n,9). In this central region the correlation of β(n,9) and MZ ψ(n,9) is 0.51, while when the KD hydropathicity scale is used for Ψ(n,9), the correlation drops to 0.37. Since the MZ and KD scales are 85% correlated [14], this shows that both MZ hydropathicity and exposed β strand propensity are central to p53 functionality. We can go further and compare the positions of the centers of the nine highest peaks of β(n,9) with the centers of the crystal core β strands [11,12]. We find excellent agreement (within one or two sites) for eight peaks, while the ninth, at 171 near the protein center, is replaced by a stabilizing α helix (Table I).

The tetramerization of p53 produces a flexible, four-armed starfish [17], quite distinct from the globular structures which most proteins (even when oligomerized) form. Thus one should not be surprised by the improved description of p53 profiles with the exposed β strand propensity scale.

The functionality of p53 as a tumor suppressor apparently depends on oligomer formation, such as tetramerization, which is driven by its core (DNA binding) domain and its tetramerization domain 325-356 near the C terminal [21,22]. At the physiological temperature of 37 degrees C, wild-type p53 is more than 50% unstructured [23]. Many studies have examined sequence-dependent p53 oligomerization mechanisms similar to amyloid beta aggregation. The most popular is the user-friendly TANGO software, which relies on the physico-chemical principles of β-sheet formation, extended by the assumption that the core regions of an aggregate are fully buried [24]. Using TANGO [24] found an epitopic β strand sequence 251-257, which is normally buried in the hydrophobic DNA binding domain (Fig. 1). Mutations that would expose this conserved ILTIITL 7-mer would promote β sheet formation and disrupt oligomer formation.

We compare our profiling results for the oligomerizing TANGO domain 251-257 with the MZ and FTI scales in Fig. 3.

Profiles of the tetramerization domain for β(n,9) and Ψ(n,9), shown in Fig. 4, exhibit the greater resolution of the FTI β scale, and can be compared to the TANGO result in Fig. 3(b) of [24].

These results suggest an epitopic model for the anomalously universal p53 tumor suppressor mechanism. Autoantibodies are recruited by p53 through real or virtual β strand epitope binding to antibody paratopes. This binding is weak, but it may be identifiable by exploiting the economy of β(n,9) and Ψ(n,9) superscale profiles, with both showing sensitive binding peaks but the former showing enhanced structure. This model can be tested against the epitopes identified by [8].

**3. Results** 15-mer overlapping epitopes numbered XY, containing amino acids 5(10X +Y) – 4 to 5(10X +Y) + 10, were winnowed by comparing patients with colon cancer to healthy controls, eventually reducing 78 starting epitopes to 11 sensitive candidate epitopes. Of these, 4 were found to be ECD successes: 9,10,25,78. These are profiled in Figs. 5-7. In some cases one sees MZ ψ(n,9) hydrophobic peaks that agree with the ECD successes, but not in other cases. In all cases there is an excellent match between the ECD successes and narrow β(n,9) peaks, usually only 7-mers. The profiles show that overlapping 15-mer epitopes -9 and -10 were successful because they share a common 7-mer. Comparison of Fig. 5(a) with Fig. 5(b) shows how much more useful the FTI βexp scale is compared to ψ hydropathicity scales.

Generally one would not expect epitopes based on other species to be as sensitive (better peaks in β(n,9)) as human epitopes, and this is usually the case. Mouse epitopes would be expected to be inferior, and this is the case for most of p53. However, near the C terminal the mouse peaks are higher. As shown in Fig. 8, this suggests an alternative choice for the -78 epitope, which could be more sensitive with a mouse sequence. Similarly the dog sequence for the -25 epitope could be better than the human sequence (Fig. 6).

The most notable and mysterious epitope is -34 (166-180), which dominates the sensitivity of the cancer surveys A and B of [8], but has below average sensitivity for ECD (C of [8], Fig. 1). As we already noted, the ninth and missing β strand should have been centered at 172, but it has been replaced by residual α helices. Because -34 is exceptional, its enlarged profiles are shown in Fig. 9, with a figure caption that may explain -34's exceptional behavior. Also Fig. 10 analyzes the marginally sensitive -43,-44,-45 cluster and suggests a possibly better choice of epitopes, which could achieve sensitivity comparable to -25 and -78.

The significance of virtual β strand epitopic binding is illustrated by the crystal structure of MDM2 binding to the transactivation domain 1-44 of p53 [26]. There have been >6000 studies of the strong and important p53-MDM2 interaction, with a view to inhibiting it to restore p53 levels [27]. The segment of p53 bound to MDM2 in the structural study is a 15-mer, 15-29, with structure identified for 17-29 [26]. The corresponding profiles are shown in Fig. 11. Although not directly useful for ECD, it is interesting that that the p53-MDM2 binding depends on large hydrogen bonding density, which is accurately recognized by FTI β transforms.

**4. Discussion** In several respects p53 can be described as a highly adaptive "scale free" hub in the self-organized cellular network [28,29], which supports the superior performance of the MZ Ψ scale compared to the KD Ψ scale. Disruption of p53's functions promotes most cancers. With respect to the new data of [8] for p53 epitopic binding to autoantibody paratopes, we have reached three main conclusions.

(1) Epitope-paratope p53 interactions are dominated by both hydropathic Ψ forces and by hydrogen bonding and Van der Waals β strand forces. Of the many proteins we have studied, the β effects are much larger for soft, starfish-shaped p53 tetramers than for typical rigid proteins like lysozyme c, where hydropathic ψ forces dominate [18], or compact tetrameric hemoglobin, which is nearly tetrahedral in shape. The qualitative importance of hydrogen bonding and Van der Waals forces for lock-and-key binding was recognized by Pauling in the 1940s, and is stated qualitatively, for example, in the study of the Fv fragment of the antibody D1.3 in its complex with rigid hen egg white lysozyme [9,30]. The present analysis using Ψ and β scaling is quantitative, and represents a substantial refinement of cellular network models of p53 [27].

(2) Our profiling p53 with hydropathic Ψ and β strand bioinformatic scales differs from hydrogen-bridged contact models [9] because it includes long range (W = 9) water network forces. Our method is much simpler and probably more accurate for p53 epitopes than methods based on all-atom force fields, where many geometrical details of paratope-epitope interactions are unknown. It is advantageous for ECD because it is economical and its only adjustable sliding window (superscaled) W parameters have a simple interpretation. [8] used overlapping 15-mers because one supposes that epitope-paratope binding is dominated by short sequences, which have here have been trimmed to be predominantly p53 7-mers. In any case, while the studies of [8] were confined to ECD of colon cancers, there are many other cancers that could be studied using similar methods. Such studies could be made more economical with the present analysis. It not only sharpens the definitions of 15-mer epitopes down to 7-mers, but it also shows (Fig. 2) that the overall number of candidate epitopes is smaller than the winnowed results of [8] could have suggested.

(3) An example of current practice is the study of liver cancer, which used a panel of 41 recombinant tumor-associated antigen fragments (obtained by ELISA) to distinguish between healthy patients and patients with diagnosable tumors [31]. This panel was reduced to 10 antigens and was as effective (40% sensitivity, 20% false positives) as other diagnostic methods, such as ultrasound. One of the antigens was a large p53 fragment, which has been shown to be much less sensitive than p53 epitopes for colon cancer [8]. It appears that aggressive cancers like liver cancer could be profitably studied with epitopic scanning of profiles from patients before and after surgery. Instead of studying the emergence of cancer in a few patients in a large trial, one could study the reverse disappearance of tumor – specific autoantibodies after a successful resection. In this case, the 50,000 patients of [8] could be replaced by 50 patients, with each patient's loss of autoantibodies followed on a suitable time scale. We suppose that this would be weeks, as aggressive cancers are expected to induce maximal effort to generate an immunological response to a severe stress, and absence of that stress could cause a rapid relaxation.

(4) It is instructive to compare the evolution of the numbers of papers on Web of Science on cancer and T cells with cancer and autoantibodies. The former tripled from 2004 to 2005, and since has increased by 50%. The latter number is about ten times smaller than the former in 2014, and has yet to break out. The former has hit its stride, with impressive results for melanoma [32]. The problem of finding pathways between 10 million cancer mutations [12,33] can be simplified by studying 6 million samples for each of 21 tumor types [34]. By using p53 epitopes and studying reversed autoantibody dynamics in liver transplant patients, one can avoid million-fold genomic questions entirely. The time may be ripe for developing p53 epitopes for early cancer screening for many cancers to break out [35].

**5. Acknowledgments** The calculations described here are very simple, given the scaling tables of [13,14] and [15]. They are most easily done on an EXCEL macro. The one used in this paper was built by Niels Voohoeve and refined by Douglass C. Allan. I have benefitted from conversations with Lawrence Williams, Martin Yarmush, and James Markmann.

| βexp9 peak site | 1TUP β strand |
|---|---|
| 123 | 123-127 |
| 143 | 141-147 |
| 159 | 156-163 |
| 172 | helix |
| 199 | 195-198 |
| 214 | 214-219 |
| 233 | 230-237 |
| 252 | 251-259 |
| 271 | 264-271 |

Table I Comparison of positions of βexp9 peak sites with β strands observed in a 96-308 p53 fragment complexed with DNA [10]. The peaks agree well with the strands, and are usually near the centers of the strands.

Figure Legends

Fig. 1. The hydroprofiles Ψ(n,W), with W = 9 and 13 exhibit abrupt edges for the 102-292 DNA binding domain of p53 [11]. The double arrows I and II are two deepest level (doubled) hydrophilic extrema for W = 9 which can synchronize mechanically for conformational changes. The synchronization principles underlying level sets are discussed in [20]. Briefly, protein conformational changes occur over long times of order μs to ms, and involve relative motions of the weakly interacting water-protein interface extrema over allometric (long) distances along the protein chain (for I here, about 60 aa, which is >> W ~ 9-13). The oligomer-forming (tetramerization) domain 325-356 also has nearly level sharp edges, which should facilitate oligomer formation. The green dashes mark the hydroneutral average for the Ψ(n,W) scales used here. Above the green lines is hydrophobic, and below is hydrophilic. The leading biomarker epitopes of [8] are also indicated.

Fig. 2. Profiles of p53 using the FTI βex Ψ scale peak more strongly than the hydropathicity MZ Ψ scale for the DNA binding [11] and tetramerization [21] domains, and the remaining peaks of the DNA BD also match the β strands found in the crystal [11], see text and Table I. The FTI scale values [15] have been multiplied by 150 to display them in the same range as the MZ scale from [14].

Fig. 3. The key oligomerizing ILTIITL 7-mer, 251-257 [24], appears as twice a stronger peak with the β strand propensity FTI scale than with the MZ scale. These profiles are shown at high resolution with W = 5. This figure can be compared to the TANGO result shown in Suppl. Fig. 3(b) of [24]. The relatively larger success of the FTI scale in finding the ILTIITL 7-mer is based on Ile, which has the second-largest value of β strand propensity on the FTI scale.

Fig. 4. The tetramerization domain 325-356 has a composite β – α structure [21]. Here W = 9 has been used to clarify the α - β separation, which is clearer with the β strand propensity scale.

Fig. 5(a). Profiles for four scales in the key -9-10 region 41-60, with W = 5. The choice of a small W produces noisy profiles, but the success of the βex scale compared to the other scales in identifying the central 7-mer 50-56 IEQWFTE epitope is clear (double arrow). Note the Trp53 = W53 at the center of this epitope, and how use of the βex scale lifts the 50-56 compressed epitope well above noise level.

Fig. 5(b). Here we have chosen W = 9, in order to reduce noise, and only the two best scales are displayed for clarity. The two best-performing epitopes, -9, -10, are associated with a single 7-mer 49-55, DIEQWFT, which is common to -9 and -10 (red double arrow). The lean goat (high fiber diet) 7-mer EDVVTWL is the only epitope studied that gives a higher peak (two V's, largest FTI βexp) than DIEQWFT, and it should be tested.

Fig. 6. The -25 epitope contains the 120-126 7-mer KSVTCTY with a βexp peak value of 206 at site 123. It appears that this 7-mer would also produce a signal for -24, but none was seen [4]. However, at 120 there are both a hydrophilic minimum (elastic hinge) and a turn in the p53 core structure bound to DNA [11]. Thus -24 would not fit stably to an antibody paratope. The corresponding dog sequence is KSVTWTY, which could be better, as βexp(W124) > βexp(C124).

Fig. 7. The -78 epitope, a 381-393 (p53 end) 13-mer, could have been more sensitive had it contained the 377-387 mer TSRHKKLMFKT. Because of proximity to the 393 end, W = 7 has been used here. The monkey sequence TSRHKKFMFKT might be better, as βexp(F383) > βexp(L383).

Fig. 8. Comparison of human and mouse W = 7 profiles (human numbering) near C terminal. It is possible that an epitope centered on 379Val (mouse numbering), such as the 7-mer 376KTMVKKV382 (red double arrow) could be a better epitopic probe than the 7-mer human epitope (blue double arrow) also shown in Fig. 7.

Fig. 9. The -34 epitope contains the 169-176 8-mer MTEVVRRC with a peak value of Ψβex9 = 227 at site 172. Comparison with Figs. 2 and 6 shows that epitope -34 has twice as large a drop of βex9 between 172 and 187 as epitope -25 has between 123 and 131. Initially the immunogenic interactions may be weaker and longer ranged and include this larger drop, so that

-34 does not form β strand bonds to antibody paratopes, whereas -25 does. At later cancer stages, the interactions could be stronger and shorter ranged, when -34, with its 8-mer edges at Ψβex9 = 200, would produce a much stronger signal than -25, with its 7-mer edges at Ψβex9 = 175.

Fig. 10. -43 to-45, shown here by three double arrows, were marginally sensitive [4]. These three 15-mers included two deep minima in β strand propensity. It is possible that a shorter 7-mer epitope, such as 213-219, would have been as successful as the four selected 15-mer epitopes 9,10,25 and 78.

Fig. 11. The structure of 15-29 (double blue arrow) a 15-mer epitope of p53 bound to a 109-residue fragment of MDM2, was studied in [27], where a short induced helical segment 19-23 (double red arrow) was found. Note that the as-grown 15-mer epitope of p53 matches the peak in virtual β strand propensity very well, suggesting that the p53-MDM2 binding depends on large hydrogen bonding density. For the MZ profile we used W = 3, even though this gives large oscillations, because the induced helical segment is short. Note that the ends of both double arrows are nearly level, and are examples of the "level set" principle [20].

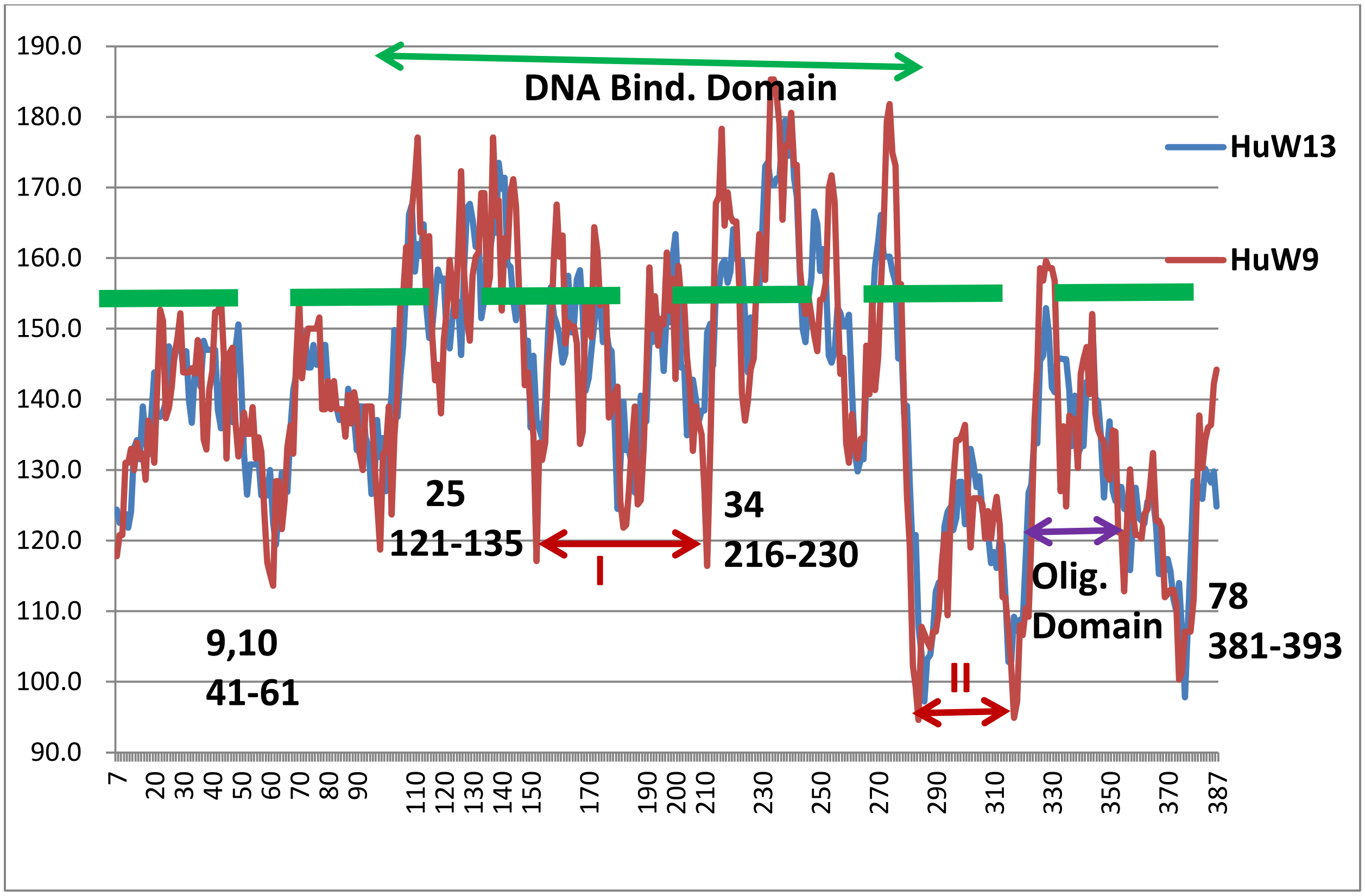

Fig. 1

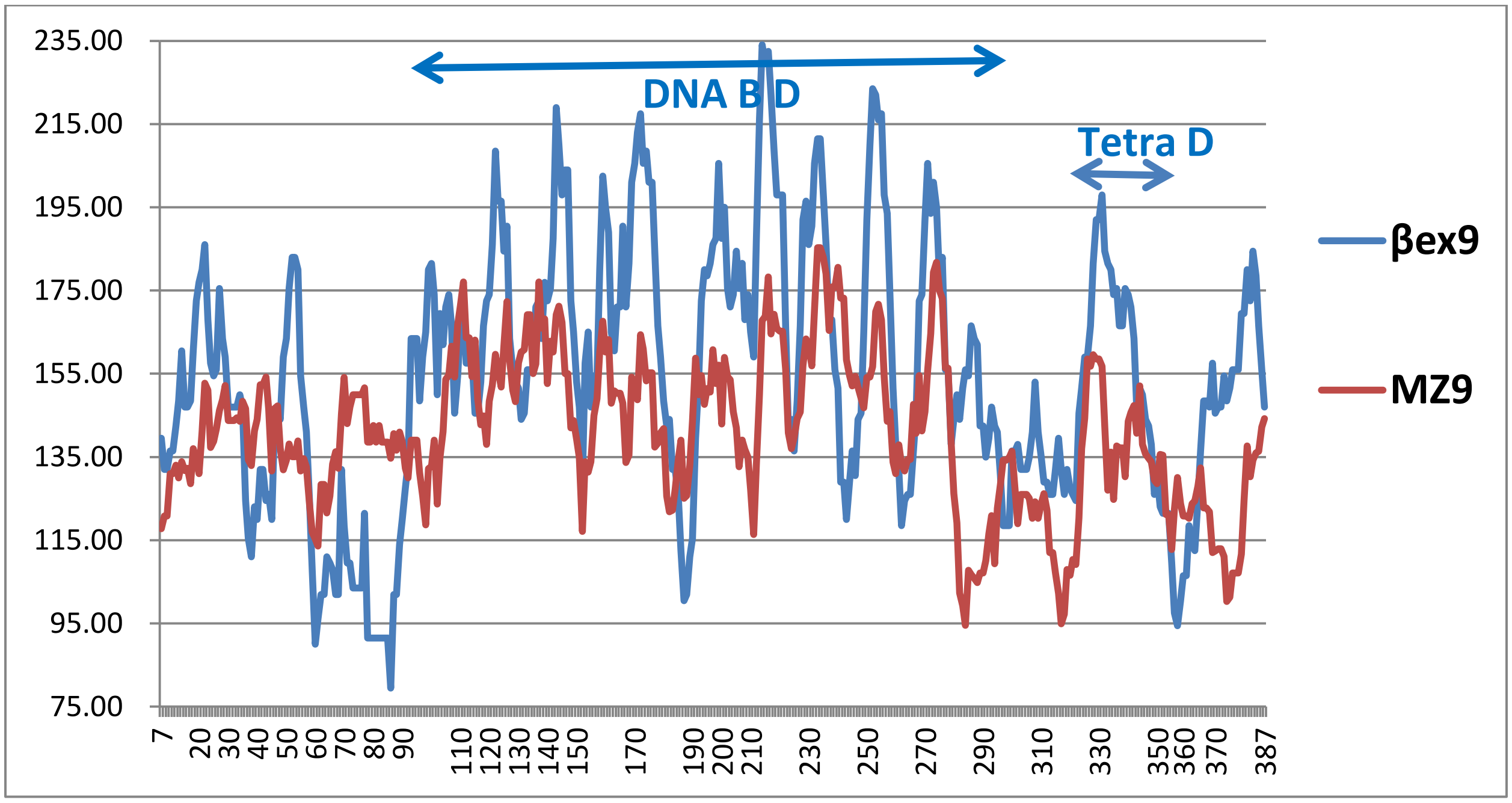

Fig. 2

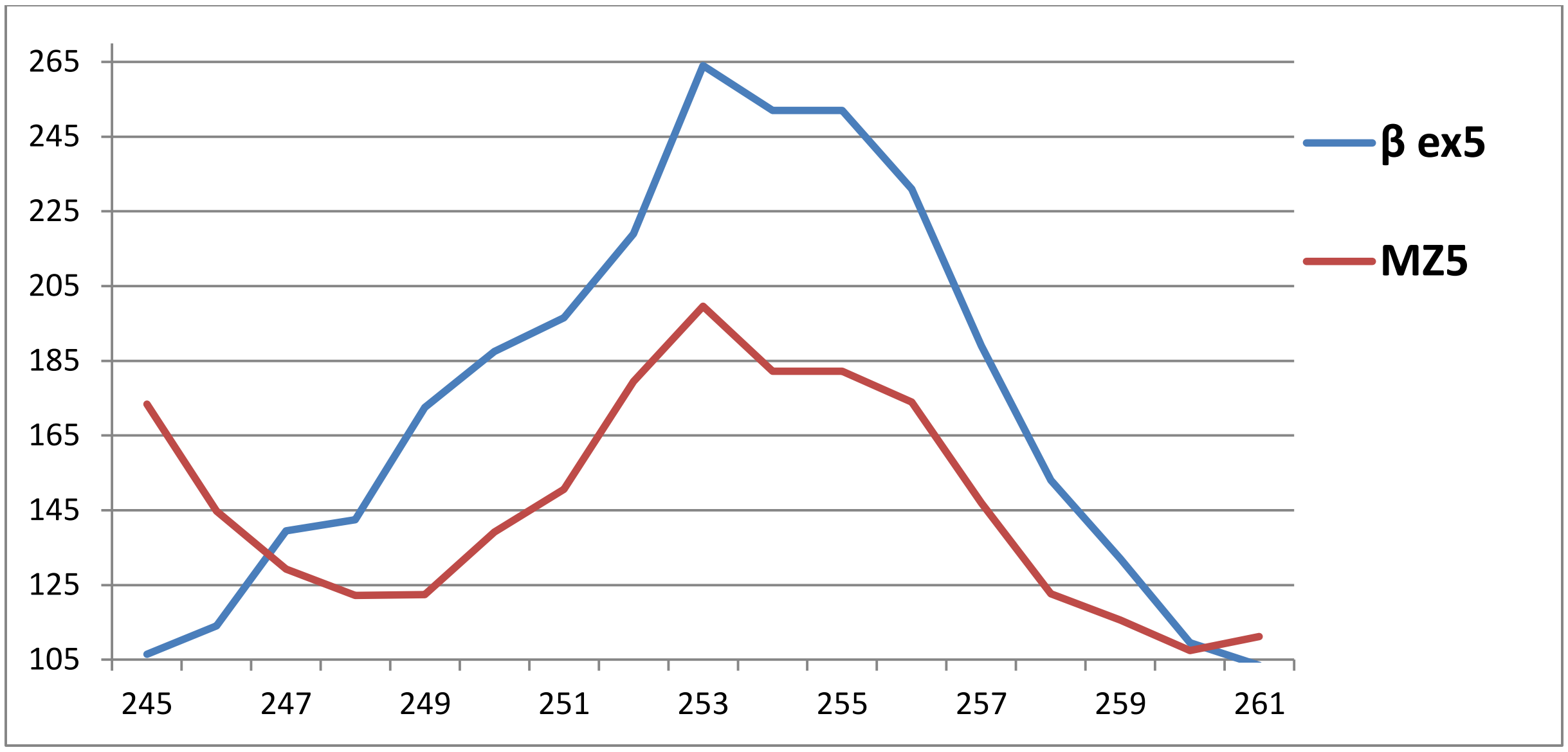

Fig. 3

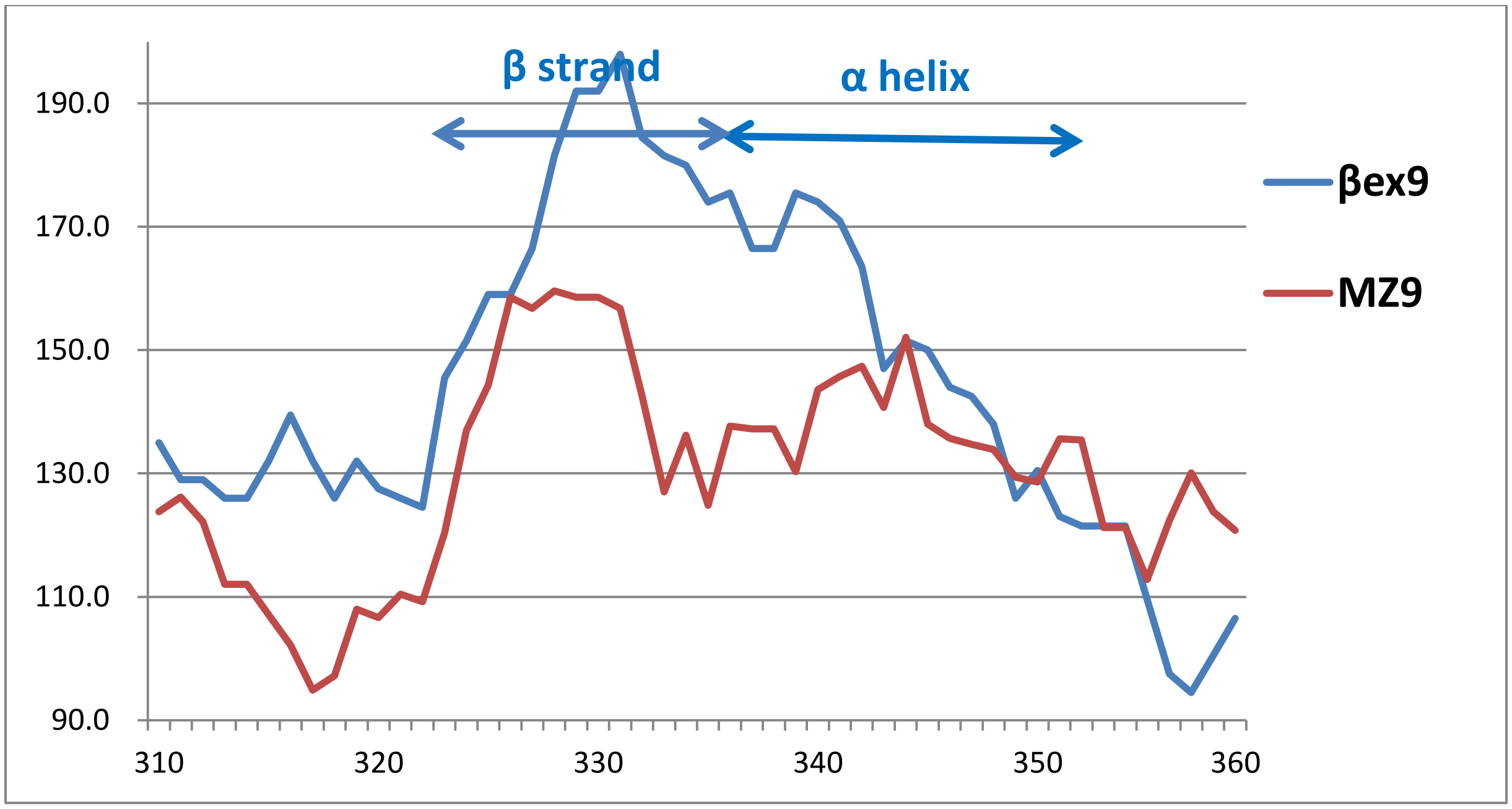

Fig. 4

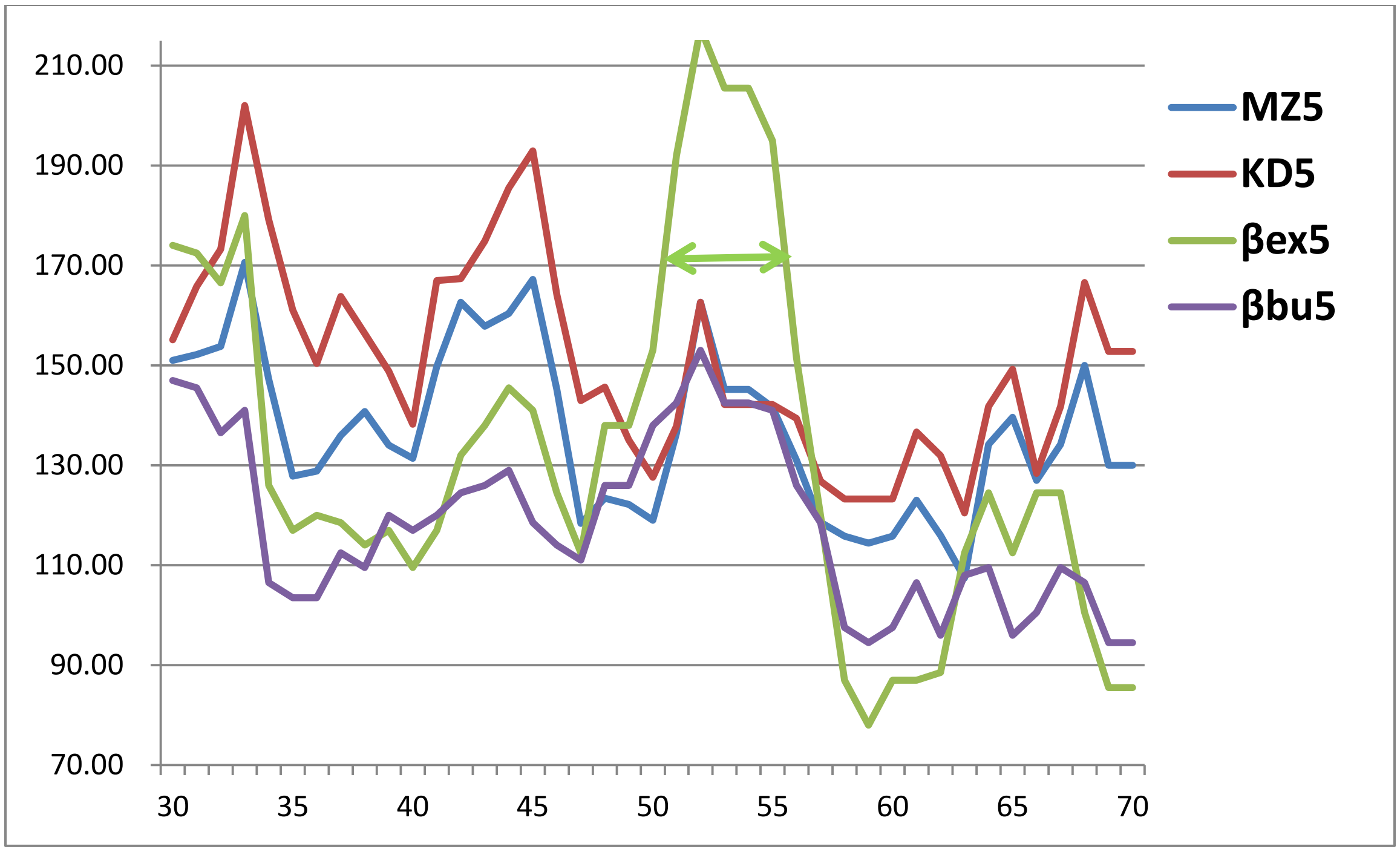

Fig. 5(a)

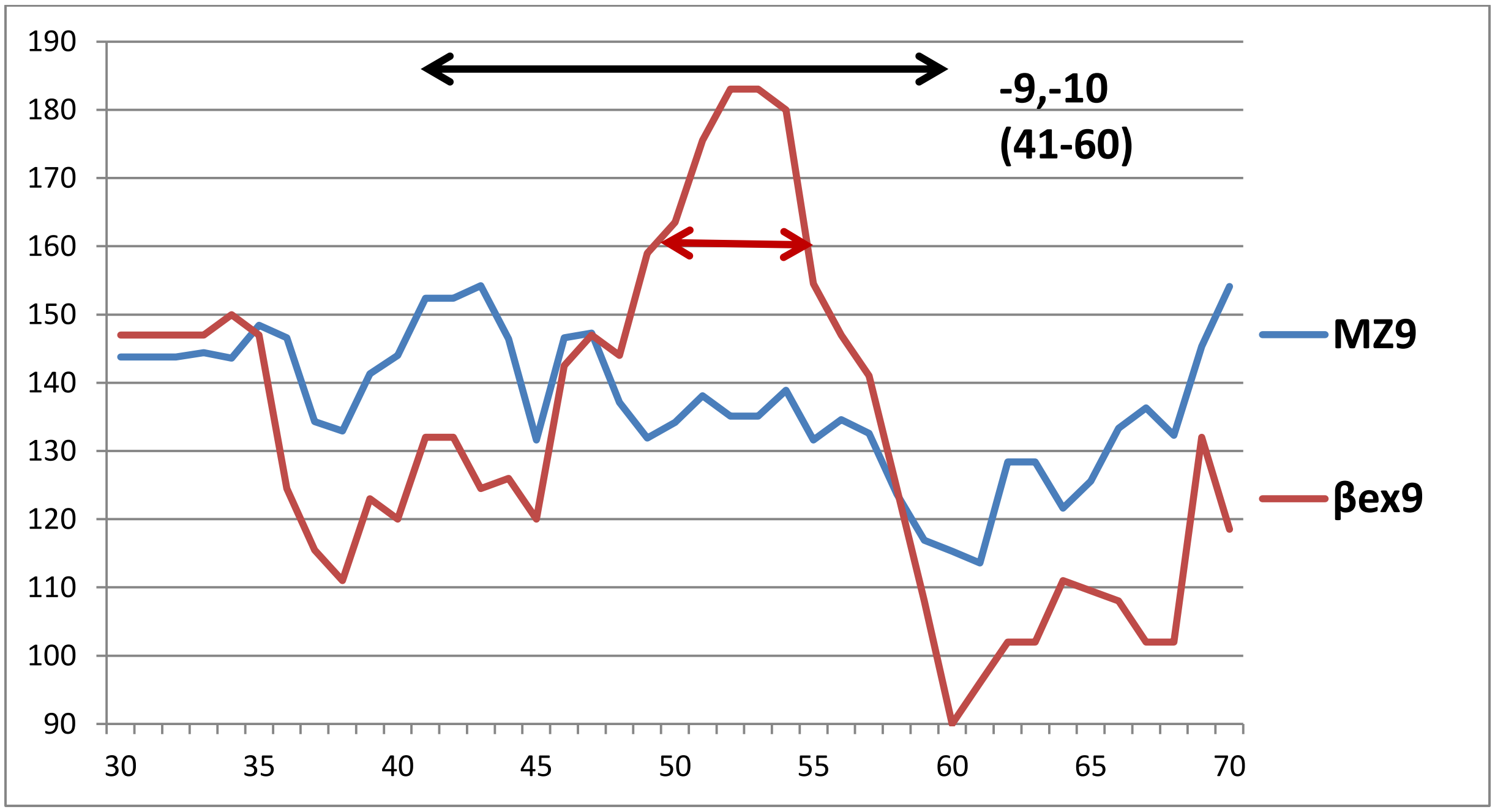

Fig. 5(b)

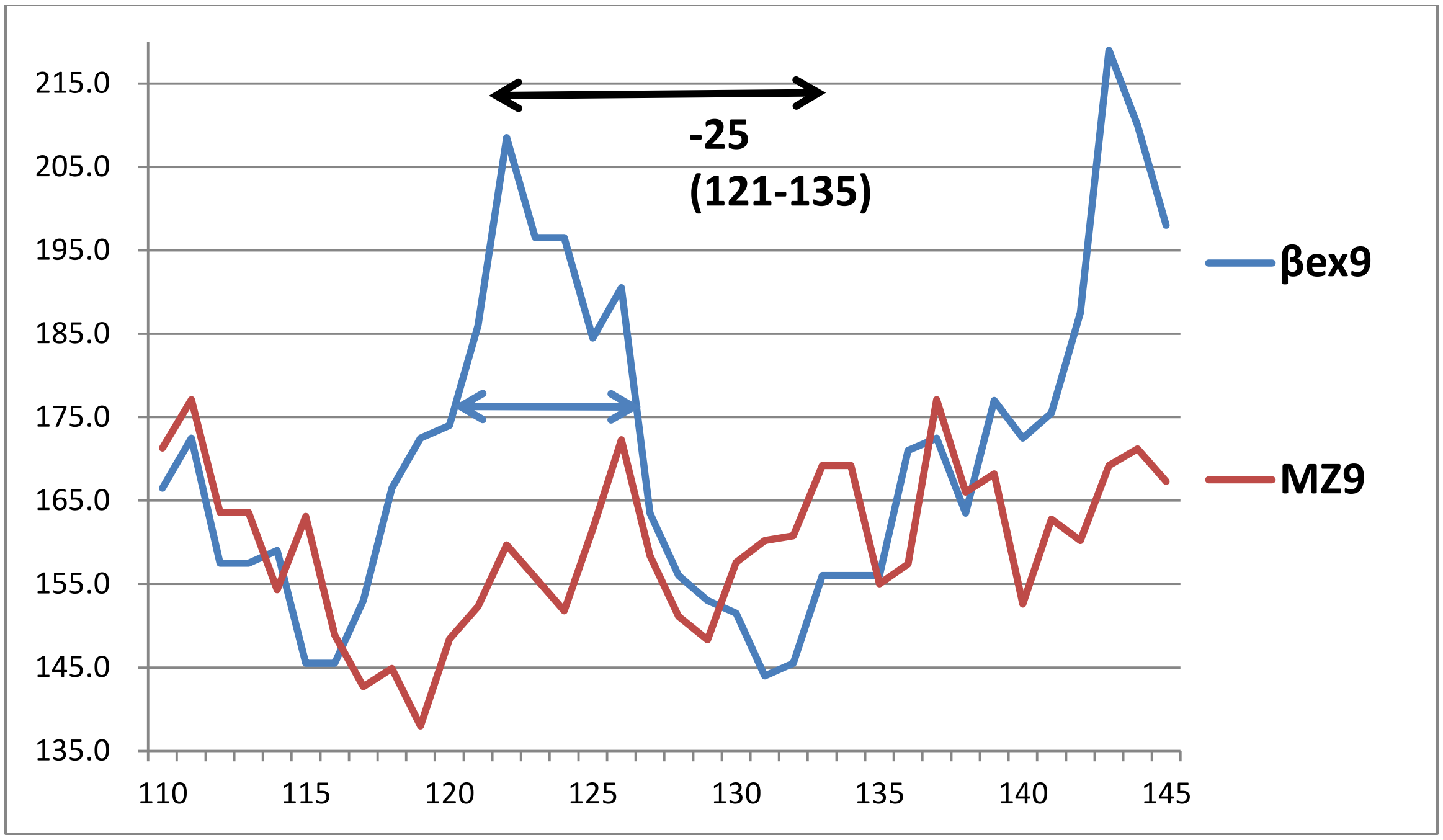

Fig. 6

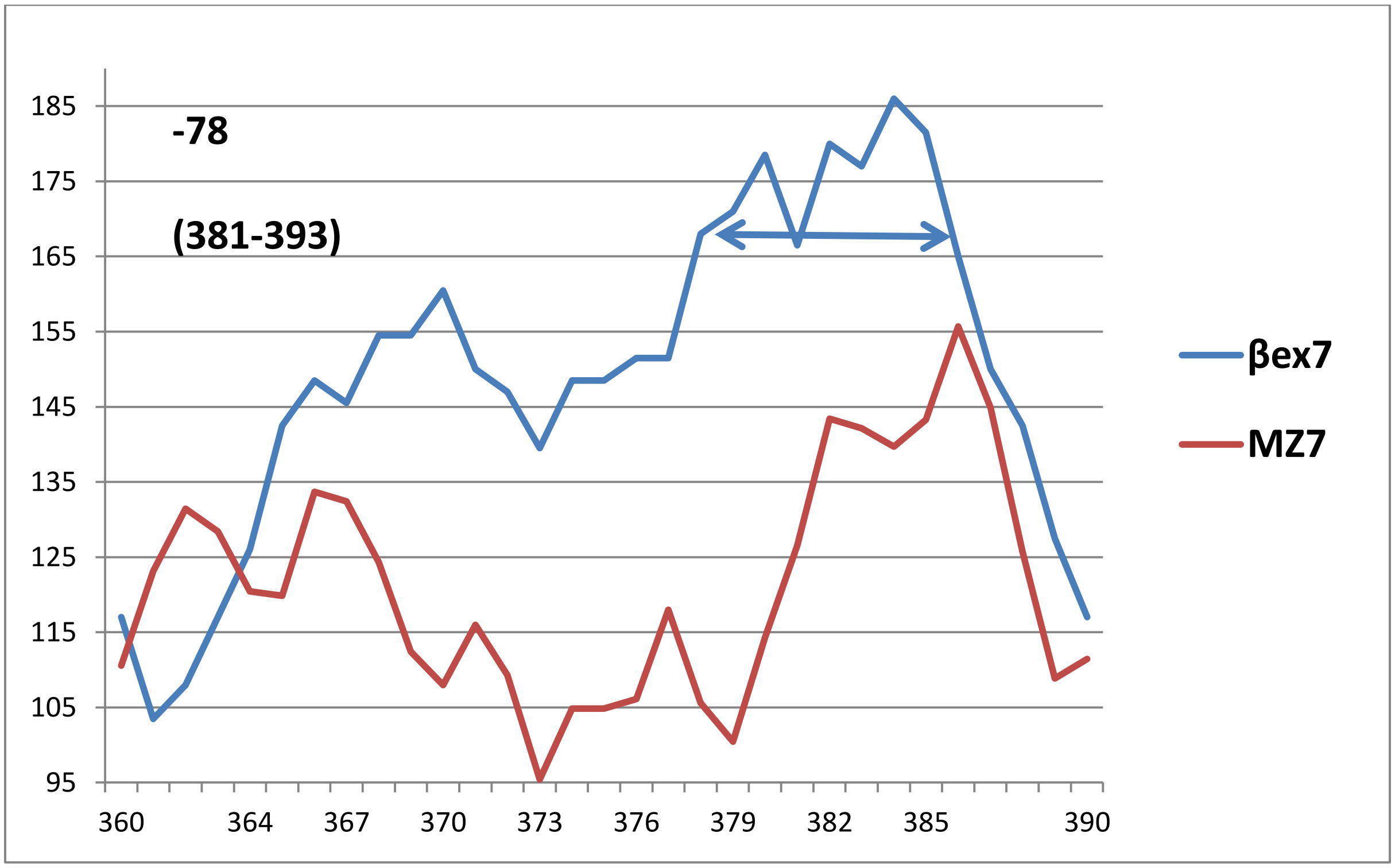

Fig. 7.

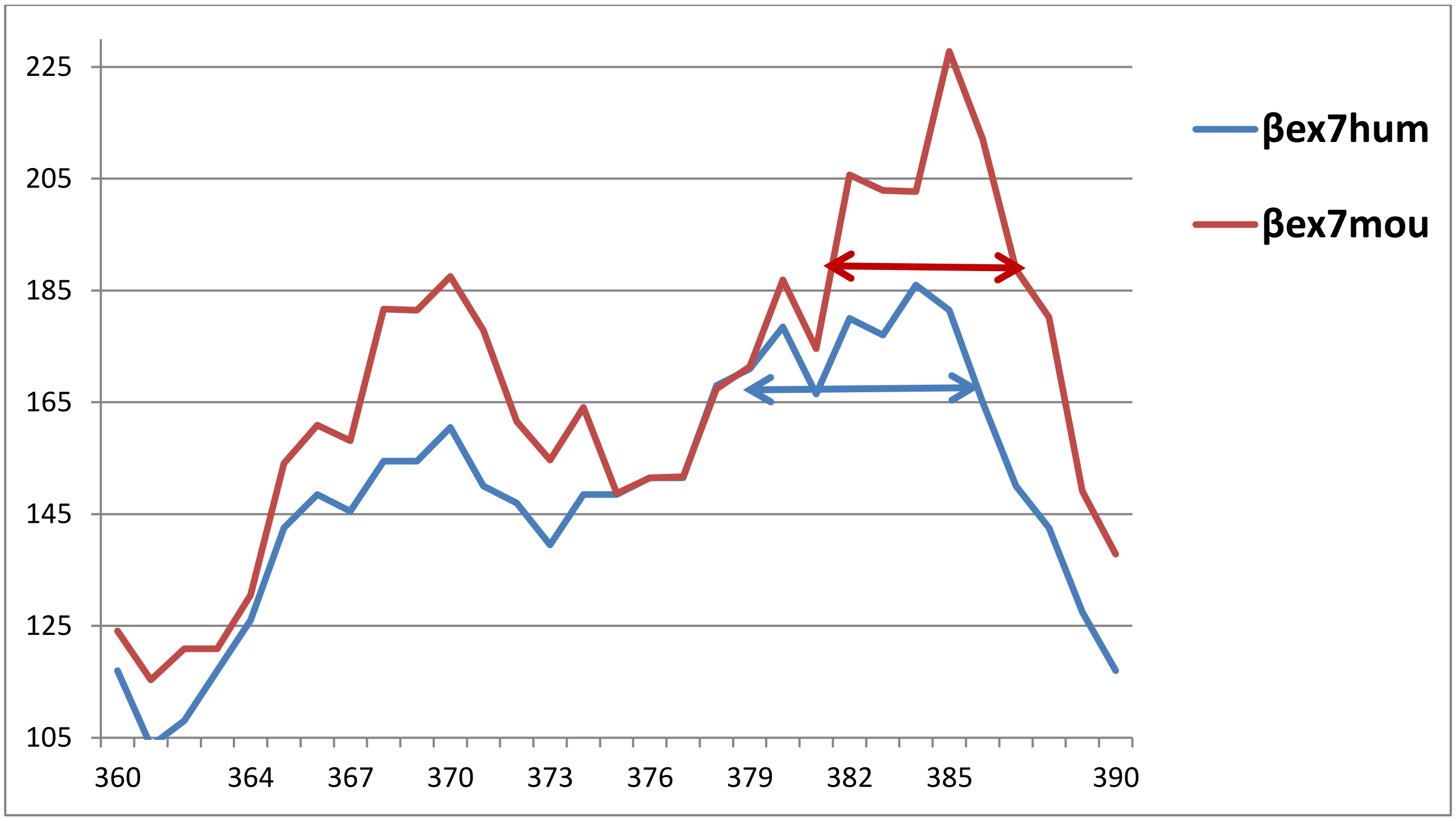

Fig. 8.

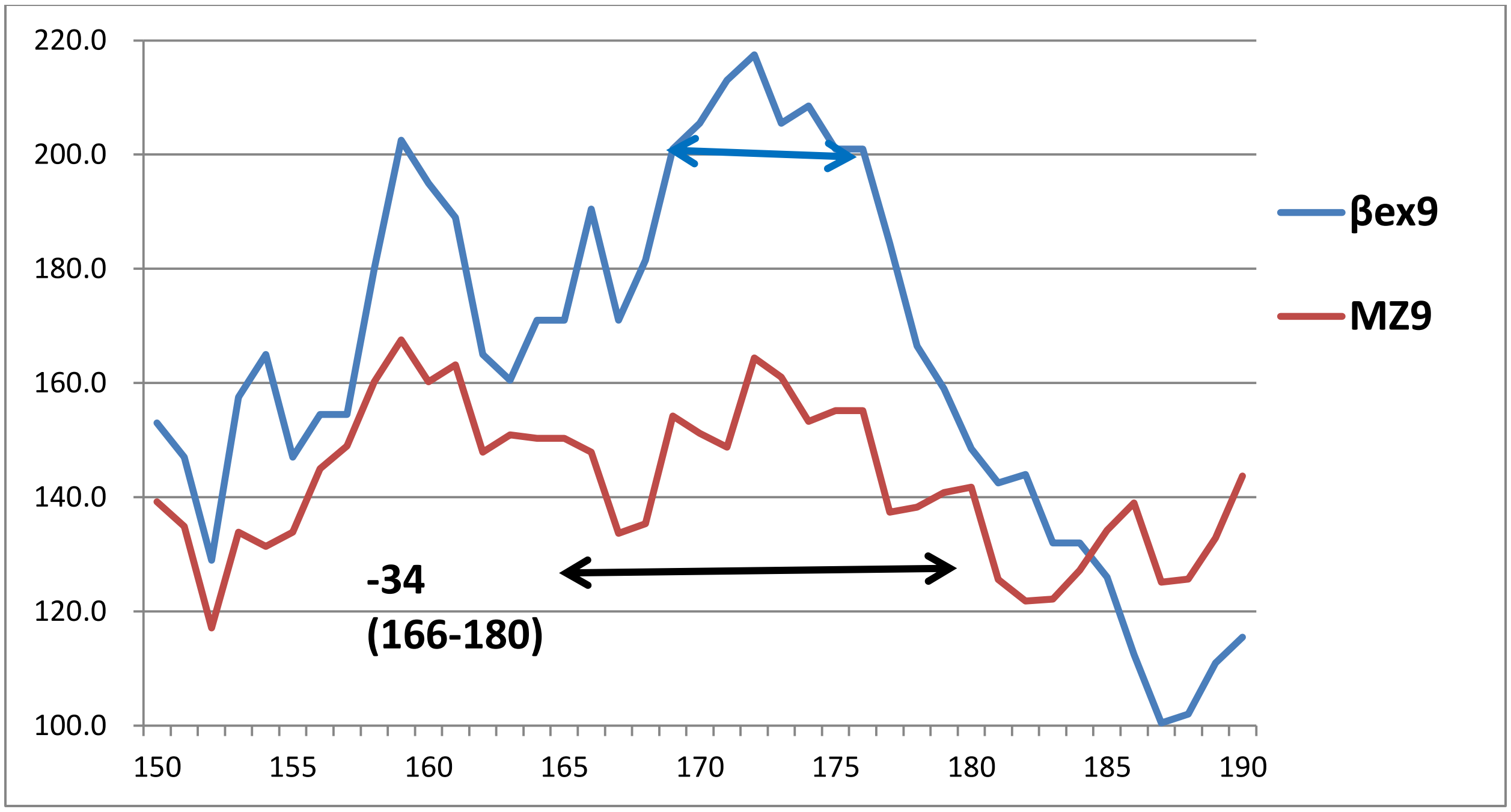

Fig. 9

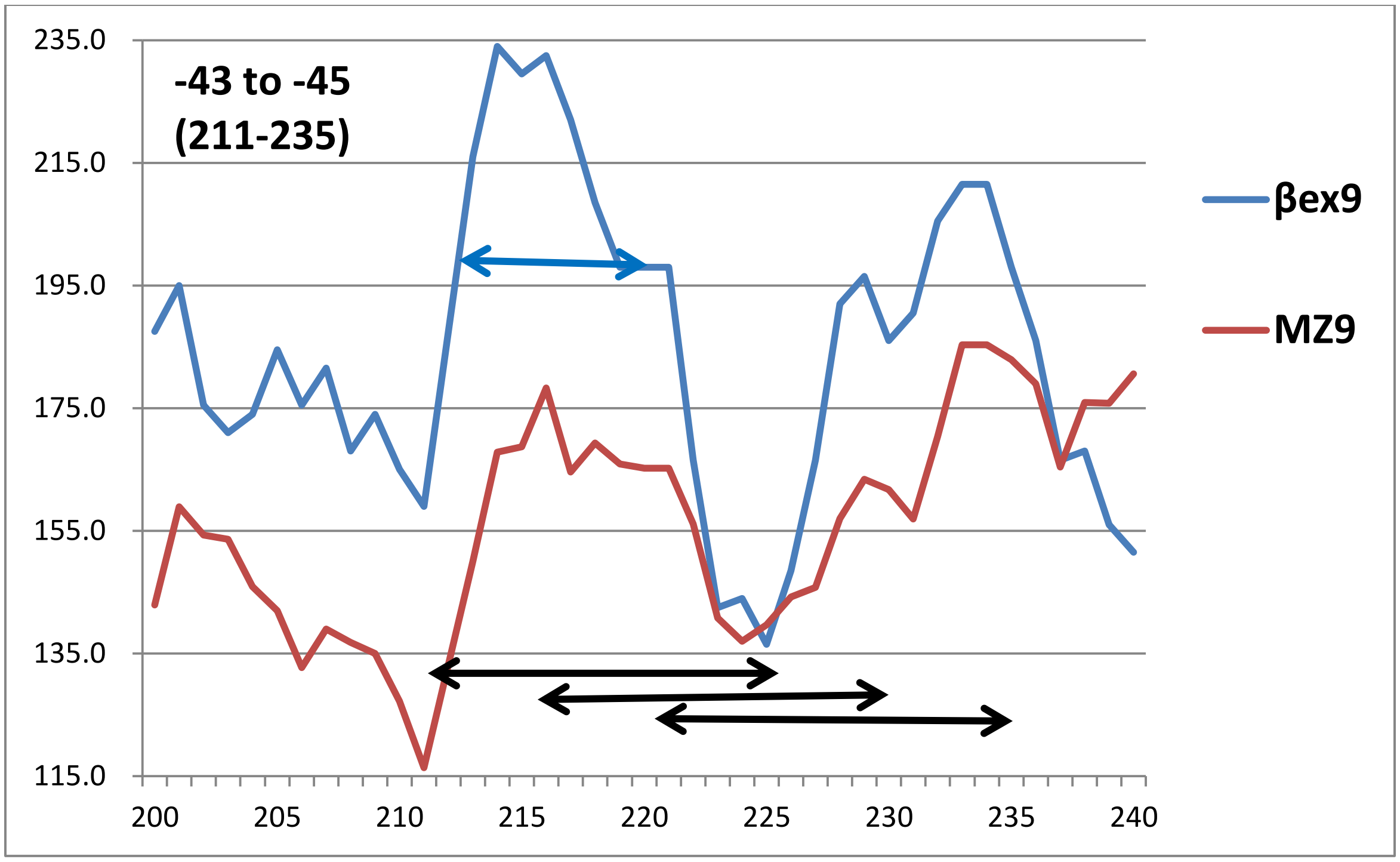

Fig. 10

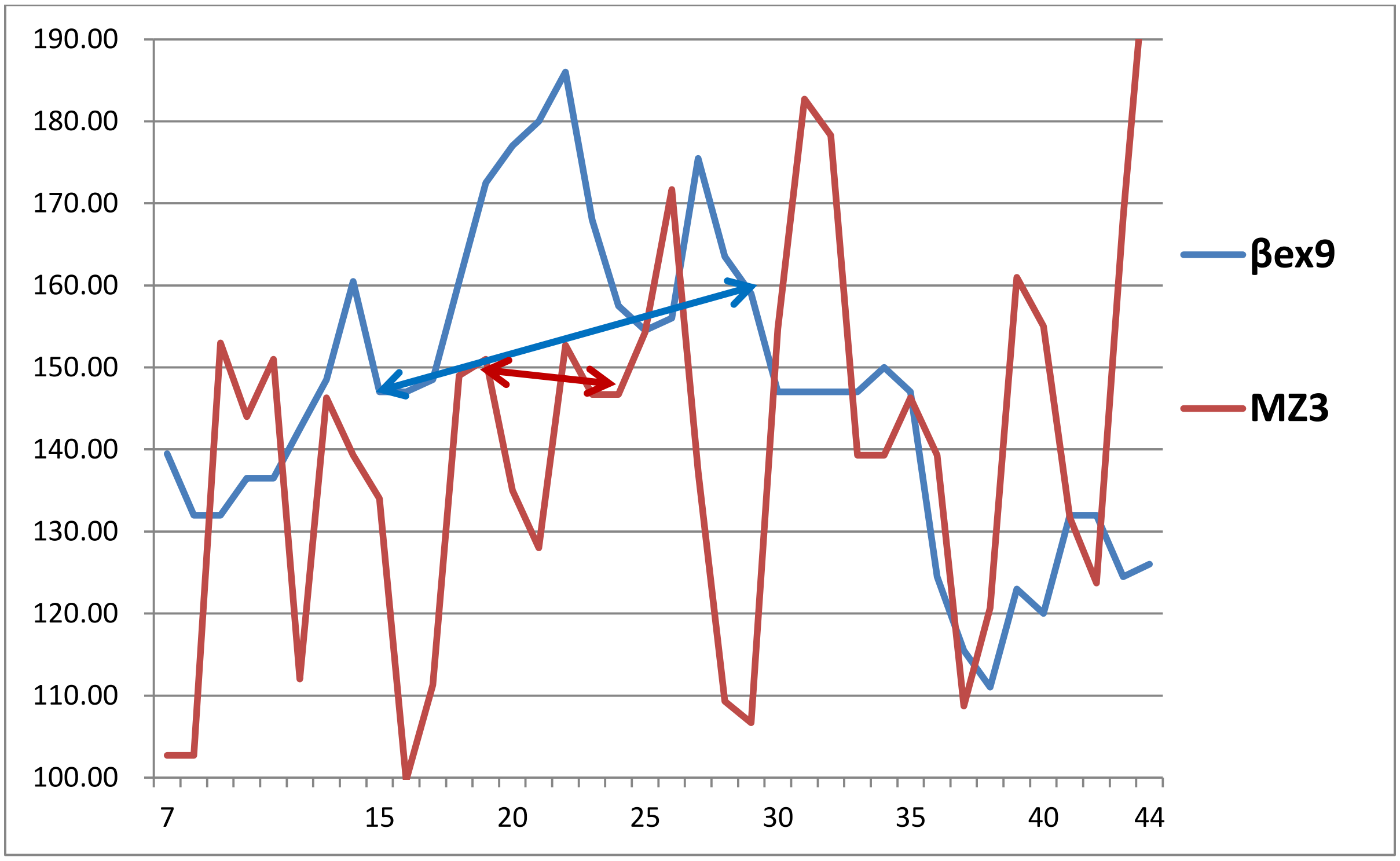

Fig. 11.